\documentstyle[12pt]{article}
\begin{document}
\begin{titlepage}
\title{{\bf Higgs boson bounds in non-minimal supersymmetric standard
models.}
\thanks{Work partly
supported by CICYT under contract AEN90-0139.}}
\vspace{8cm}
\author{{\bf J. R. Espinosa} and {\bf M. Quir\'os} \thanks{Talk
presented at the XXVI INTERNATIONAL CONFERENCE ON HIGH ENERGY
PHYSICS, August 6-12, 1992, Dallas, Texas.}\\
Instituto de Estructura de la Materia\\
Serrano 123, E-28006 Madrid. Spain}
\date{}
\maketitle
\def\baselinestretch{1.2}

\begin{abstract}
{\large\noindent In the minimal supersymmetric standard model (MSSM), when
radiative corrections are included, the mass of the
$CP=+1$ lightest Higgs
boson is bounded by $\sim 110\ GeV$ for $m_t < 150\ GeV$ and a
scale of supersymmetry breaking $\sim\ 1\ TeV$. In non-minimal
supersymmetric standard models (NMSSM) upper bounds on the mass of
the corresponding scalar
Higgs boson arise if the theory is required to remain
perturbative up to scales $\gg G_F^{-1/2}$. We have computed those
bounds for two illustrative NMSSM: i) A model with an arbitrary
number of gauge singlets; ii) A model with three $SU(2)_L$ triplets
with $Y=0,\pm 1$. We have integrated numerically the corresponding
renormalization  group {\mbox equations} (RGE), including the top and bottom
quark Yukawa couplings, and added one-loop radiative corrections. For
$m_t > 91\ GeV$ the absolute bounds are $\sim 140\ GeV$ for both
models.}
\end{abstract}
\vskip-15cm
\rightline{{\bf IEM-FT-60/92}}
\rightline{{\bf August 1992}}
\end{titlepage}
\newpage
\def\baselinestretch{1.3}
\large
\section{Introduction}

One of the most constraining features of the MSSM is the existence of
an absolute upper bound on the tree-level mass of the lightest
scalar ($CP=+1$) Higgs boson
\begin{equation}
m_h \leq m_Z\mid \cos 2\beta\mid,
\end{equation}
where $\tan \beta \equiv v_2/v_1$, $v_i\equiv <H_i^o>$. Therefore a negative
result in the Higgs search at LEP-200
$^1$ would seem to exclude the {\mbox MSSM} making its search at NLC/LHC/SSC
unnecessary. However, before ruling out phenomenological supersymmetry
we should consider all possible effects spoiling the relation
(1):
radiative corrections and the introduction of models with an
enlarged Higgs sector. As
we will see both effects allow to overcome the bound (1) but
nevertheless there are still very constraining bounds in general
models.

\section{Radiative Corrections}

They have been computed by {\mbox different} groups using different methods:
standard diagrammatic techniques $^2$, the one-loop effective
potential $^3$ and the RGE approach $^{4,5}$. The latter approach is
reliable provided that $\Lambda^2_S/m^2_W \gg 1$, where $\Lambda_S$ is
the scale of supersymmetry breaking, since it amounts to a
resummation of all leading logarithms in the effective potential.

Radiative corrections depend on $\tan \beta$, $m_t$ and $\Lambda_S$.
In the MSSM the absolute upper bound on $m_h$, for $\mid \cos 2\beta
\mid =1$, is modified by radiative corrections to a $\Lambda_S$
and $m_t$-dependent bound. For $\Lambda_S= 1\ TeV$ we have
parametrized it as $^5$:
\begin{equation}
m_h(GeV)= 111+80x+34x^2,
\end{equation}
where $x=m_t/150\ GeV -1$.

Radiative corrections can be sizeable, and even
larger than the tree-level mass, thus
putting doubts on the validity of the perturbative expansion. The RGE
approach has been used to study the stability of one-loop versus
two-loop corrections $^5$. It has been estimated analytically, and
checked numerically, that
\begin{equation}
\frac{\Delta^{2-loop}m_h^2}{\Delta^{1-loop}m_h^2}\sim
-\ 0.1 \frac{\alpha}{\pi \sin^2\theta_W}
\left(\frac{m_t}{m_W}\right)^2,
\end{equation}
which guaranties the validity of perturbative expansion.
{}From eq. (2) we see that for $m_t<150\
GeV$, $m_h<111\ GeV$ which is the range allowed by radiative
corrections in the MSSM.

\section{Non-minimal Supersymmetric Standard Models}

The tree-level bound (1) does no longer hold in supersymmetric models
with extra Higgs fields, $i. e.$ in NMSSM. We will study in this
section Higgs bounds in a general class of NMSSM.

The first (obvious) enlargement of the Higgs sector consists in adding
pairs of Higgs doublets $H_1^{(j)}$, $H_2^{(j)}$, $j=1,...,N$. These
{\mbox models} have been analyzed $^6$
and their lightest {\mbox sca-}lar Higgs boson shown to
have the tree-level bound (1).

Consider now NMSSM with Higgs doublets
$H_1$, $H_2$ and neutral scalar fields
$N_{12}^{(i)},\ N_{11}^{(j)}$,\newline
$\ N_{22}^{(j)}$ (either $SU(2)_L
\times U(1)_Y$ {\mbox sin-}glets or making part of higher dimensional
representations) with a cubic superpotential $f=g+f_{YUK}$
\begin{equation}
g= \vec{\lambda}\cdot\vec{N}_{12}H_1^o H_2^o +
\sum_{i=1}^2 \vec{\chi}_{i}\cdot\vec{N}_{ii}(H_i^o)^2 ,
\end{equation}
where $\vec{\lambda}\cdot\vec{N}\equiv
\sum_j \lambda_j N^{(j)}$ and $f_{YUK}$ contains all Yukawa
couplings giving mass to fermions.
Then, the lightest scalar Higgs boson mass has an upper bound given by $^7$
\begin{equation}
\displaystyle{\frac{m_h^2}{v^2}}\leq \frac{1}{2}(g^2+g'^2)\cos^22\beta
+\vec{\lambda}^2\sin^22\beta
+\vec{\chi}_{1}^2\cos^4\beta +
\vec{\chi}_{2}^2\sin^4\beta,
\end{equation}
where $g,\ g'$ are the $SU(2)_L\times U(1)_Y$ couplings and
$v^2\equiv v_1^2+v_2^2$.

The bound for the MSSM is recovered from (5) when
$\vec{\lambda}=\vec{\chi}_{1}=\vec{\chi}_{2}=0$.
However in {\mbox NMSSM} some of the Yukawa couplings
in (4) can be non-zero. In that case the upper bound on the
lightest scalar Higgs boson mass comes from the requirement that the
supersymmetric theory remains perturbative up to some scale
$\Lambda$, in the energy range where the theory holds.

We will keep in $f_{YUK}$ the top and bottom quark Yukawa couplings, $i. e.$
\begin{equation}
f_{YUK}=h_t  Q\cdot H_2 U^c +h_b  Q\cdot H_1 D^c,
\end{equation}
with boundary conditions
\begin{equation}
\begin{array}{cc}
h_t={\displaystyle\frac{g}{\sqrt{2}}\frac{m_t}{m_W}}(1+\cot^2\beta)^{1/2},&
h_b={\displaystyle\frac{g}{\sqrt{2}}\frac{m_b}{m_W}}(1+\tan^2\beta)^{1/2}.
\end{array}
\end{equation}
$m_t$ in (7) will be considered as a variable while $h_b$ is fixed by
$m_b$, which is taken to be $m_b(2\ m_b)=\ 5\ GeV$. For $\tan
\beta\gg 1$, $h_b$ can become important. In particular it is
comparable to $h_t$ for $\tan\beta\sim m_t/m_b$. $h_{\tau}$ will be
neglected since it is given by $h_b (m_{\tau}/m_b)$ for all values
of $\tan\beta$. The cubic $g$-superpotential in (4) and so the
tree-level mass in (5) are model dependent. The latter depends on the
couplings $\vec{\lambda}$, $\vec{\chi}_i$
allowed by the perturbative requirement.

Radiative corrections will be included using the RGE approach $^4$.
This procedure is universal in the sense of assuming that the {\mbox standard}
model holds below $\Lambda_S$ and the supersymmetric theory beyond
$\Lambda_S$. We have taken here $\Lambda_S=1\ TeV$. The radiative mass
$\Delta m^2_r(\beta)$ is $\beta$-dependent and has to be added to
(5). In the following we will consider two generic {\mbox NMSSM}.

\subsection{NMSSM with an arbitrary number of singlets}

These models are defined by a Higgs sector containing $H_1$, $H_2$
and $n$ singlets $S_i\ (i=1,...,n)$ with a cubic superpotential
\begin{equation}
g= \vec{\lambda}\cdot \vec{S} H_1\cdot H_2 + \frac{1}{6} \sum_{i,j,k}
\chi_{ijk} S_iS_jS_k.
\end{equation}
The model with $n=1$ has been studied in great detail in the
literature $^{6-9}$. The tree-level upper bound on the mass of the
lightest scalar Higgs boson for the case of arbitrary $n$ can be written
as$^{7,10}$:
\begin{equation}
m_h^2\leq {\displaystyle\left(\right.} \cos^2 2\beta
+ {\displaystyle\frac{2\vec{\lambda}^2
\cos^2\theta_W}{g^2}}\sin^22\beta \left.\right)m_Z^2.
\end{equation}
The relevant one-loop RGE are
\[
\begin{array}{c}
4\pi^2\dot{\vec{\lambda}^2}=\left\{ -\frac{3}{2}g^2 - \frac{1}{2}g'^2
+2\vec{\lambda}^2  + \frac{3}{2}(h_t^2 + h_b^2)\right\}\vec{\lambda}^2
+ \frac{1}{4}\lambda_i\lambda_j tr(M_iM_j),\vspace{.4cm}\\
8\pi^2\dot M_k=3\lambda_k \vec{M}\cdot\vec{\lambda} + \frac{3}{4}
tr(\vec{M}\cdot M_k)\cdot \vec{M},\vspace{.4cm}\\
8\pi^2\dot h_t=\left\{ -\frac{3}{2}g^2 - \frac{13}{18}g'^2
- \frac{8}{3}g_s^2
  +\frac{1}{2}\vec{\lambda}^2+ 3h_t^2 + \frac{1}{2} h_b^2\right\}h_t,
\vspace{.4cm}\\
8\pi^2\dot h_b=\left\{ -\frac{3}{2}g^2 - \frac{7}{18}g'^2
- \frac{8}{3}g_s^2
  +\frac{1}{2}\vec{\lambda}^2+ \frac{1}{2}h_t^2
+ 3 h_b^2\right\}h_b,\vspace{.4cm}\\
\end{array}
\]
\begin{equation}
\begin{array}{c}
16\pi^2\dot g=g^3,\vspace{.4cm}\\
16\pi^2\dot g'=11 g'^3,\vspace{.4cm}\\
16\pi^2\dot g_s=-3g_s^3,
\end{array}
\end{equation}
where $g_s$ is the $SU(3)$
gauge coupling and $(M_k)_{ij}\equiv \chi_{ijk}$. The key observation
to maximize $\vec{\lambda}^2$, and so the bound (9), is the property
$\lambda_i \lambda_j tr(M_iM_j)\geq 0$ which follows trivially from
the definition of $M_j$.

Assuming that the theory remains perturbative up to the scale
$\Lambda= 10^{17}\ GeV$, integrating numerically the RGE and including
radiative corrections for $\Lambda_S=1\ TeV$ we find $^{10}$ the upper
bound shown in Fig. 1 in the $(m_h,m_t)$-plane.

We see from Fig. 1
that the detailed functional dependence of $m_h$ on $m_t$ is
parametrized by the value of $\tan\beta$.
The dashed curve where the solid lines stop
correspond to values of $m_t$ such that the Yukawa coupling $h_t$
becomes non-perturbative. (For $\tan\beta>30$ the corresponding lines
would follow very close to the $\tan\beta=20$ curve in Fig. 1, but stopping at
lower values of $m_t$ because of the large values of $h_b$.) The dotted
curve on the top of the figure is the enveloping for all values of
$\tan\beta$ and can therefore be considered as the absolute upper
bound. Of course once the top quark will be discovered, and its mass
known, the bound on $m_h$, and its
$\tan\beta$-dependence, will become more dramatic. For instance, for
$\tan\beta\gg 1$ the bound becomes undistinguishable with that in the
MSSM. Using the constraints $\tan\beta\geq 1$ and $m_t\geq 91\ GeV$
we obtain from Fig. 1, $m_h\leq 140\ GeV$.

\subsection{NMSSM with $Y=0,\pm 1$ $SU(2)_L$ triplets}

This model can be considered as an example where
$\vec{\chi}_{i}\neq 0\ (i=1,2)$ in (4). It is the supersymmetric
extension of a non-supersymmetric standard model with Higgs
triplets $^{11}$ which do not break the custodial symmetry at the
tree-level provided there is a particular relation between the vacuum
expectation values of their neutral components.

The Higgs content is $H_1,\ H_2$ and $\Sigma$, $\Psi_1$, $\Psi_2$,
which are $SU(2)_L$ triplets with hypercharges $0,\ \pm 1$, respectively,
field content
\begin{equation}
\begin{array}{c}
\Sigma=\left(\begin{array}{cc}
\xi^o/\sqrt{2}&-\xi_2^+\\
\xi_1^-&-\xi^o/\sqrt{2}
\end{array}\right),\vspace{.4cm}\\
\Psi_1=\left(\begin{array}{cc}
\psi_1^+/\sqrt{2}&-\psi_1^{++}\\
\psi_1^o&-\psi_1^+/\sqrt{2}
\end{array}\right),\;
\Psi_2=\left(\begin{array}{cc}
\psi_2^-/\sqrt{2}&-\psi_2^o\\
\psi_2^{--}&-\psi_2^-/\sqrt{2}
\end{array}\right),
\end{array}
\end{equation}
and a cubic superpotential
\begin{equation}
g=\lambda_1 H_1\cdot \Sigma H_2 + \lambda_2 tr\Sigma\Psi_1\Psi_2
+\chi_1 H_1\cdot \Psi_1 H_1 + \chi_2 H_2\cdot \Psi_2 H_2.
\end{equation}
The tree-level bound on the mass of the lightest scalar Higgs boson can be
written as $^7$:
\begin{equation}
m_h^2\leq \left\{
\begin{array}{c}
  \\
\end{array}
 \cos^2 2\beta + \left[\lambda_1^2\sin^22\beta
+2(\chi_1^2\cos^4\beta+ \chi_2^2\sin^4\beta)\right]
{\displaystyle\frac{\cos^2\theta_W}{g^2}}\right\}m_Z^2.
\end{equation}
The one-loop RGE are:
\begin{equation}
\begin{array}{c}
8\pi^2\dot\lambda_1=\left\{ -\frac{7}{2}g^2 - \frac{1}{2}g'^2
+2\lambda_1^2 + \frac{1}{2}\lambda_2^2
+ 3 \chi_1^2+3 \chi_2^2 + \frac{3}{2}(h_t^2+h_b^2)\right\}\lambda_1,
\vspace{.4cm}\\
8\pi^2\dot\lambda_2=\left\{-6g^2 - 2 g'^2 +\frac{1}{2}\lambda_1^2
+\frac{3}{2} \lambda_2^2 + \chi_1^2+ \chi_2^2\right\}
\lambda_2,\vspace{.4cm}\\
8\pi^2\dot\chi_1=\left\{ -\frac{7}{2}g^2 - \frac{3}{2}g'^2
+\frac{3}{2}\lambda_1^2 + \frac{1}{2}\lambda_2^2
+ 7\chi_1^2 + 3h_b^2\right\}\chi_1,
\vspace{.4cm}\\
8\pi^2\dot\chi_2=\left\{ -\frac{7}{2}g^2 - \frac{3}{2}g'^2
+\frac{3}{2}\lambda_1^2 + \frac{1}{2}\lambda_2^2
+ 7\chi_2^2 + 3h_t^2\right\}\chi_2,\vspace{.4cm}\\
8\pi^2\dot h_t=\left\{ -\frac{3}{2}g^2 - \frac{13}{18}g'^2
- \frac{8}{3}g_s^2
+\frac{3}{4}\lambda_1^2 + 3\chi_2^2 + 3h_t^2 +\frac{1}{2} h_b^2\right\}h_t,
\vspace{.4cm}\\
8\pi^2\dot h_b=\left\{ -\frac{3}{2}g^2 - \frac{7}{18}g'^2
- \frac{8}{3}g_s^2
+\frac{3}{4}\lambda_1^2 + 3\chi_1^2 + 3h_b^2 +\frac{1}{2} h_t^2\right\}h_b,
\vspace{.4cm}\\
16\pi^2\dot g=7g^3,\vspace{.4cm}\\
16\pi^2\dot g'=17g'^3,\vspace{.4cm}\\
16\pi^2\dot g_s=-3g_s^3.
\end{array}
\end{equation}
We have integrated numerically the RGE assuming that the theory
remains perturbative up to the scale $\Lambda = 10^{14}\ GeV$ (the scale
where the gauge coupling constants become non-perturbative) and included
the radiative corrections for $\Lambda_S = 1\ TeV$.
We find $^{10}$
the upper bound on $m_h$ as a function of $m_t$  for
$\tan \beta \le 20$ in Fig. 2, and for
$\tan \beta > 20$ in Fig. 3. The dashed curves correspond again
to the region where $h_t$ becomes non-perturbative. We see that the
maximum upper bound, $m_h\sim 140\ GeV$, corresponds to values of
$\tan\beta$ much larger than one. For $\tan\beta>50$ $h_b$
becomes non-perturbative, and the corresponding curves would fall, very
rapidly with increasing $m_t$.
\newpage
\def\baselinestretch{1.2}
\large
\section*{Figure Captions}
\begin{description}
\item[Fig. 1] Upper
bounds on the lightest scalar Higgs boson in NMSSM with singlets.
\item[Fig. 2] Upper
bounds on the lightest scalar Higss boson in NMSSM with $Y=0,\pm 1$
triplets for \(\tan\beta\leq 20\).
\item[Fig. 3] Upper
bounds on the lightest scalar Higss boson in NMSSM with $Y=0,\pm 1$
triplets for \(\tan\beta\geq 30\).
\end{description}
\end{document}